

\input phyzzx

\catcode`@=11
\newtoks\KUNS
\newtoks\HETH
\newtoks\monthyear
\Pubnum={KUNS~\the\KUNS\cr HE(TH)~\the\HETH}
\monthyear={November, 1992}
\def\p@bblock{\begingroup \tabskip=\hsize minus \hsize
   \baselineskip=1.5\ht\strutbox \topspace-2\baselineskip
   \halign to\hsize{\strut ##\hfil\tabskip=0pt\crcr
   \the\Pubnum\cr hep-ph/9210229\cr \the\monthyear\cr }\endgroup}
\def\bftitlestyle#1{\par\begingroup \titleparagraphs
     \iftwelv@\fourteenpoint\else\twelvepoint\fi
   \noindent {\bf #1}\par\endgroup }
\def\title#1{\vskip\frontpageskip \bftitlestyle{#1} \vskip\headskip }
%
%
\def\Kyoto{\address{Department of Physics,~Kyoto University \break
                            Kyoto~606,~JAPAN}}

%
%
\paperfootline={\hss\iffrontpage\else\ifp@genum%
                \tenrm --\thinspace\folio\thinspace --\hss\fi\fi}
\footline=\paperfootline
%
%

%
\def\journal#1&#2(#3){\begingroup \let\journal=\dummyj@urnal
    \unskip, \sl #1\unskip~\bf\ignorespaces #2\rm
    (\afterassignment\j@ur \count255=#3) \endgroup\ignorespaces }
\def\andjournal#1&#2(#3){\begingroup \let\journal=\dummyj@urnal
    \sl #1\unskip~\bf\ignorespaces #2\rm
    (\afterassignment\j@ur \count255=#3) \endgroup\ignorespaces }
\def\andvol&#1(#2){\begingroup \let\journal=\dummyj@urnal
    \bf\ignorespaces #1\rm
    (\afterassignment\j@ur \count255=#2) \endgroup\ignorespaces }
\def\NP{Nucl.~Phys. }
\def\PR{Phys.~Rev. }
\def\PRL{Phys.~Rev.~Lett. }
\def\PL{Phys.~Lett. }

%
%
%
\def\ee{\eqno\eq }
\KUNS={1162}     
\HETH={92/12}   

\def\1#1{{1 \over {#1}}}
\def\2#1{{{#1} \over 2}}

\def\F{{\rm F}}
\def\B{{\rm B}}

\def\calD{{\cal D}}

\def\tm{{\widetilde m}}
\def\tl{{\widetilde \lambda}}
\def\th{{\widetilde h}}

\def\smg{{\raise2.2pt\hbox{$>$}}\kern-8pt\lower3pt\hbox{$\sim$}}
\def\simg{\ \smg \ }
\def\sml{{\raise2.2pt\hbox{$<$}}\kern-7pt\lower3pt\hbox{$\sim$}}
\def\siml{\ \sml \ }

\def\lbar{\overline}
\def\MS{${\lbar {\rm MS}}$\ }
\def\ep{effective potential}

\def\sm{{1 \over 2}\lambda\phi^2+m^2}
\def\fm{g^2\phi^2}
\def\der#1{{\partial \over \partial #1}}
\def\SUM#1#2{\sum_{\scriptstyle #1
                   \atop \lower2pt\hbox{$\scriptstyle #2$}}}

\catcode`@=12

\def\widebar#1{\vbox{\ialign{##\crcr
          \hskip 1.0pt\hrulefill\hskip 1.0pt
          \crcr\noalign{\kern-1pt\vskip0.07cm\nointerlineskip}
          $\hfil\displaystyle{#1}\hfil$\crcr}}}
\def\1#1{{1 \over {#1}}}
\def\2#1{{{#1} \over 2}}
\def\NL{\nolimits}

\def\F{{\rm F}}
\def\B{{\rm B}}

\def\MB{M_\B}
\def\MF{M_\F}
\def\calD{{\cal D}}

\def\tm{{\widetilde m}}
\def\tl{{\widetilde \lambda}}
\def\th{{\widetilde h}}
\def\tg{{\widetilde g}}
\def\tphi{{\widetilde \phi}}
\def\ts{{\widetilde s}}
\def\tMF{{\widetilde \MF}}
\def\tbeta{{\widetilde \beta_{g1}}}

\def\simg{\gsim}
\def\siml{\lsim}

\def\lbar{\overline}
\def\MS{${\lbar {\rm MS}}$\ }
\def\ep{effective potential}

\def\sm{{1 \over 2}\lambda\phi^2+m^2}
\def\fm{g^2\phi^2}
\def\der#1{{\partial \over \partial #1}}
\def\SUM#1#2{\sum_{\scriptstyle #1
                   \atop \lower2pt\hbox{$\scriptstyle #2$}}}
\def\region{g^2\phi^2 \simg m^2}
\def\cregion{g^2\phi^2 \siml m^2}
\REF\BKMNI{
   M.~Bando, T.~Kugo, N.~Maekawa and H.~Nakano,
   ``Improving the Effective Potential", Kyoto preprint KUNS 1161,
   September, 1992. }
\REF\AC{T.~Appelquist and J.~Carazzone    \journal \PR &D11 (75) 2856;
 \nextline See also, K.~Symanzik \journal Comm. Math. Phys. &34 (73) 7.
}
\REF\KY{
   Y.~Kazama and Y.-P.~Yao   \journal \PRL &43 (79) 1562;
   \andjournal \PR &D21 (80) 1116; \andvol &D21(80)1138;
   \andvol &D25 (82) 1605;
   \nextline   C.K.~Lee   \journal \NP &B161 (79) 171;
   \nextline   T.~Hagiwara and N.~Nakazawa
      \journal \PR &D23 (81) 959.}
\REF\Wein{
   S.~Weinberg  \journal \PL &91B (80) 51;
   \nextline B.~Ovrut and H.~Schnitzer \journal \PR &D21 (80) 3369;
   \andvol &D22 (80) 2518; \andjournal \NP &B179 (81) 381;
   \nextline P.~Bin{\'e}truy and T.~Sch{\"u}cker
   \journal \NP &B178 (81) 293; \andvol &B178(81)307;
   \nextline H.~Georgi and S.~Dawson \journal \NP &B179 (81) 477.}
\REF\Georgi{
   H.~Georgi \journal \NP &B361 (91) 51. }
\REF\CW{
   S.~Coleman and E.~Weinberg \journal Phys. Rev. &D7 (73) 1888;
      \nextline
        S.~Weinberg \journal Phys. Rev. &D7(73) 2887.}

\REF\EJ{M.~B.~Einhorn and D.~R.~T.~Jones
      \journal \NP &B230[FS10] (90) 261.}

\titlepage

\title{Improving the Effective Potential: Multi-Mass-Scale Case }

\author{   Masako~BANDO      }
\address{   Aichi University, Miyoshi, Aichi 470-02      }

\andauthor{   Taichiro~KUGO,\ \ \ Nobuhiro~MAEKAWA        }
\vglue0.1cm
\centerline{ and }
\vglue0.1mm
\titlestyle{\twelvecp    Hiroaki~NAKANO      }

\Kyoto

\abstract{
Previously proposed procedure for improving the
effective potential by using renormalization group equation (RGE)
is generalized so as to be applicable to any system
containing several different mass scales.
If one knows $L$-loop effective potential
and $(L+1)$-loop RGE coefficient functions, this procedure gives
an improved potential which satisfies the RGE and
contains all of the leading, next-to-leading,
$\cdots $, and $L$-th-to-leading log terms. Our procedure here also
clarifies how naturally the so-called effective field theory can be
incorporated in the RGE in \MS scheme.
 }

\endpage          

\sequentialequations 

In a previous paper,\refmark{\BKMNI}
 we have presented a procedure for improving the \ep\
so as to satisfy the renormalization group equation (RGE).
By knowing  $L$-loop effective potential
and $(L+1)$-loop RGE coefficient functions,
the procedure gives an improved potential which contains all of the
leading,
next-to-leading, $\cdots $, and $L$-th-to-leading log terms.
However, its applicability was restricted to systems which possess
essentially a single mass scale.

The purpose of this paper is to generalize the procedure so as to be
applicable to any system possessing multi-mass-scales. The main idea
is to
make use of the decoupling theorem.\refmark{\AC}  By this theorem,
it is made sufficient to treat essentially a single log factor at any
scale of
field strength, since all the heavy particles (heavier than that scale)
decouple and all the light particles (lighter than that scale) yield
essentially the same log factors.
In other words, we treat effective field theory\refmark{\KY-\Georgi}
in each interval between mass thresholds,
in which heavier particles decouple and lighter particles may be
regarded as massless. So the problem of improving \ep\ reduces
to that for a single mass scale system and our previous procedure
becomes
applicable.
Our novel recognition in this context is that the RGE's
of those effective field theories, apparently different interval by
interval,
are in fact the same one.
This guarantees that we are solving the {\it same } RGE for the
\ep, using different effective field theories depending on the field
scale.

To explain our procedure, we consider the following Yukawa model which
is probably the simplest system possessing two mass scales:
$$
{\cal L} = {1\over 2}(\partial \phi )^2 -{1\over 2}m^2 \phi ^2
   -{1\over 4!}\lambda \phi ^4 + \widebar\psi (i \rlap/\partial
-g\phi )\psi  - hm^4 \ ,
\eqn\eqmodel
$$
where $\phi $ is a single component massive real scalar field and
$\psi =(\psi _1, \cdots ,\psi _N)^T$
is an $N$-component {\it massless} Dirac spinor field.
[We take the Dirac field to
be $N$-component simply because the factor $N$ may play the role of a
{\it tracer} of the fermion loop contributions.] Note that the
 masslessness
of the fermion is protected by the invariance under `chiral-parity'
transformation: $\phi  \rightarrow  -\phi , \ \psi  \rightarrow
\exp(i\gamma _5\pi /2)\psi $.
The last term $hm^4$ in the Lagrangian \eqmodel\
is the vacuum energy term which is usually omitted
but, as noted in the previous paper,\refmark{\BKMNI}
becomes relevant to us in the
calculation of the \ep\ in the mass-independent renormalization
scheme.

The \ep\  satisfies the RGE:
$$
\eqalignno{
{\cal D}\, V&(\phi ,m^2,g^2,\lambda ,h;\mu ) = 0 \ ,   &\eqname\eqRGE
  \cr
\noalign{\vskip 0.5cm}
{\cal D} = \mu \der\mu +\beta _\lambda \der\lambda &+\beta _g\der{g^2}
-\gamma _mm^2\der{m^2}-\gamma \phi \der\phi
+\beta _h\der{h} \ .             &\eqname\eqRGED \cr
}$$
The solution is well-known:
$$
V(\phi ,m^2,g^2,\lambda ,h;\ \mu ^2) =
   V\big(\widebar\phi (t),\widebar m^2(t),\widebar g^2(t),
   \widebar\lambda (t),\widebar h(t);\,\ e^{2t}\!\mu ^2\big) \ ,
\eqn\eqSOLUTION
$$
where $\widebar g^2$, $\widebar\lambda $, $\widebar m^2$,
$\widebar\phi $ and
$\widebar h$ are
running parameters whose $t$-dependence is determined by
the RG running equations
$d\widebar g^2(t)/dt = \beta _g\big(\widebar g^2(t), \widebar\lambda
(t)\big), \
d\widebar m^2(t)/dt = -\gamma _m\big(\widebar g^2(t), \widebar\lambda
(t)\big)
\,\widebar m^2(t)$, \
$d\widebar h(t)/dt = \beta _h\big(\widebar g^2(t),
\widebar\lambda (t), \widebar h(t)\big)$, and so on,
with the boundary condition that they reduce to the unbarred
parameters at
$t=0$.

The solution \eqSOLUTION\ gives full information of RGE:
\ As a result of the fact
that RGE is a first order differential equation, the \ep\
is determined once its function form is known at a certain value of
$t$.
So, to derive  useful information from RGE,
we need to know the function of \ep\ at a certain value of $t$,
a `boundary' function.

To see the logarithm structure of the \ep, let us first write
the quantum Lagrangian
in the following form by rescaling the fields by a factor $g$:
$$
{\cal L}= {1\over g^2} \bigg[
{1\over 2}\big(\partial (g\Phi )\big)^2 -{1\over 2}m^2 (g\Phi )^2
-{1\over 4!}({\lambda \over g^2})(g\Phi )^4 + (g\widebar\psi )
[i \rlap/\partial -(g\Phi )](g\psi )
- g^2hm^4 \bigg] .
\ee
$$
Then, to compute the \ep\ $V(\phi )$, we make the field shift $\Phi
\rightarrow \Phi +\phi $
and regard $g\Phi $ and $g\psi $ as our basic quantum fields.
In this form the parameters characterizing the theory are only
the scalar and fermion masses $\MB$ and $\MF$  (in the presence of
scalar background $\phi $),
$$
\MB^2\equiv {1\over 2}\lambda \phi ^2+m^2, \qquad
\MF\equiv g\phi  \ ,
\eqn\eqmass
$$
the cubic coupling $(\lambda /g^2)\MF$, the quartic coupling
$\lambda /g^2$,
and $g^2$ (aside from the vacuum-energy term). Moreover the
last parameter $g^2$ is no longer
the Yukawa coupling constant but an overall factor
in front of the action just like Planck constant $\hbar $.
Then, it is clear that the
$L$-loop level contribution to the \ep\  has the following form:
$$
V^{(L)} = (g^2)^{L-1}\MB^4 \times
\big[\hbox{function in } \ \ln{\MF^2\over \mu ^2}, \ \ln{\MB^2\over
\mu ^2},
\ {\MF^2\over \MB^2}, \ {\lambda \over g^2} \big] \ .
\eqn\eqLLOOP
$$
We have two logarithm factors $\ln(\MF^2/\mu ^2)$ and
$\ln(\MB^2/\mu ^2)$
in this two mass scale system.  For the purpose of the leading-log
series
expansion below, we express the latter
log as $\ln(\MB^2/\mu ^2) = \ln(\MF^2/\mu ^2)+\ln(\MB^2/\MF^2)$, and
introduce the following variables:
$$
\eqalign{
s &\equiv  g^2\ln{\MF^2\over \mu ^2} \ , \qquad
u \equiv  g^2\ln{\MB^2\over M_\F^2} \ , \cr
x &\equiv  {\MF^2 \over \MB^2} \ , \qquad
y \equiv  {\lambda \over  g^2} \ , \qquad  z \equiv   g^2 h{m^4\over
\MB^4} \ . \cr
}\ee
$$
Since we know that the logarithms appear only up to $L$-th power at
the $L$
loop level, the $L$-loop contribution \eqLLOOP\ takes the form
\def\vl#1{v^{(L)}_{#1}(x,y)}
$$
\vl{L,0}
V^{(L)}
= g^{-2}\MB^4  \, \sum_{\ell=0}^L \sum_{k=0}^{L-\ell} \ (g^2)^{L
-(\ell+k)}
   \, \vl{\ell,k} \, s^\ell u^k \ ,
\eqn\eqLOGEXP
$$
so that the full \ep\  has the form:
$$
\eqalign{
 V &=  g^{-2}\MB^4 \sum_{\ell=0}^\infty  \, g^{2\ell}
   [ f_\ell(s, u, x, y) + z\delta _{\ell,0} ] \ ,\cr
 f_\ell&(s, u, x, y)
   =\sum_{L=\ell}^\infty  \sum_{k=0}^{L-\ell} \,
   \vl{L-(\ell+k),k}\, s^{L-(\ell+k)}u^k \ . \cr
}\eqn\eqLLOG
$$
Just as in the previous paper,\refmark{\BKMNI}
this form of expansion \eqLLOG\ of the \ep\  in powers of $g^2$
gives a {\it leading-log series expansion}: namely, the functions
$f_0, f_1, \cdots $ correspond to the leading, next-to-leading,
$\cdots $,
log terms, respectively. So the explicit $g^2$ factors, which appear
when the expression is written in terms of variables
$s, u, x, y, z$ and $g^2$,  show the
order in this leading-log series expansion. We refer to the term
proportional
to $(g^2)^{\ell-1}$ in $V$ as $\ell$-th-to-leading log term.

The second equation in \eqLLOG\ tells us that the
$\ell$-th-to-leading
log function $f_\ell$ at $s=0$ in particular is given by
$$
f_\ell(s=0, u, x, y)
   =\sum_{L=\ell}^\infty \, \vl{0,L-\ell} \, u^{L-\ell} \ .
\ee$$
Namely, the information of the $(\ell+k)$-loop level potential
$V^{(L=\ell+k)}$ determines the $u^k$ term of the function
$f_\ell(s=0, u, x, y)$, \ie, the $(g^2)^{\ell-1}u^k$ term in
$V\big\vert_{s=0}$.
Therefore if we {\it restrict} ourselves to the region of $\phi $
in which $u$ is as small as an
$O(g^2)$ quantity (in the sense of leading-log series expansion),
\ie, to the region
$$
\ln{\MB^2\over \MF^2} \siml O(1) \quad  \rightarrow  \quad  g^2\phi ^2
 \simg m^2\ ,
\ee
$$
then the $L$-loop potential $V_L=V^{(0)}+V^{(1)}+\cdots +V^{(L)}$
at $s=0$ already gives the \ep\  {\it `exact'}
up to $L$-th-to-leading log order:
$$
V\big\vert_{s=0} = V_L\big\vert_{s=0} + O(g^{2L}) \ .
\ee
$$
That is, in such a region of $\phi $, we can use the function
$V_L\big\vert_{s=0}$
as a `boundary' function required in the RHS of
the solution \eqSOLUTION\ of RGE.
Therefore, with the $L$-loop potential $V_L$ at hand,
the \ep\ satisfying the RGE can be given by
$$
V(\phi ,m^2,g^2,\lambda ,h;\ \mu ^2) =
   V_L\big( \widebar\phi (t),\widebar m^2(t),\widebar g^2(t),\widebar
\lambda (t),\widebar h(t);
      \ e^{2t}\!\mu ^2\big)\Big\vert_{\widebar s(t)=0}\ ,
\eqn\eqSOLTWO
$$
with $\widebar s(t)$ being the $s$ variable at `time' $t$:
$$
\widebar s(t) \equiv
   \widebar g^2(t)\ln{\widebar M_{\rm F}^2(t)\over e^{2t}\mu ^2} \ ,
\quad
\widebar M_{\rm F}^2(t)\equiv  \widebar g^2(t)\widebar \phi ^2(t) \ .
\eqn\eqSBAR
$$
The barred quantities in the solution \eqSOLTWO\
should of course be evaluated at $t$ satisfying $\widebar s(t)=0$.

Our solution \eqSOLTWO\ is `exact' only up to $L$-th-to-leading log
order
and only in the region $\region$.
However, even with this approximate boundary function,
RGE is satisfied {\it exactly}
if the runnings of the barred quantities are
solved exactly, of course. To satisfy also the RGE only up to
the $L$-th-to-leading log order, it is sufficient to solve the
runnings of
the parameters $\widebar g^2/g^2, \widebar\lambda /\lambda , \widebar
\phi /\phi , \widebar m^2/m^2$ and
$\widebar h/h$  up to {\it $L$-th power in } $g^2$
in the sense of leading-log series expansion, and for this order of
accuracy,
the $(L+1)$-loop RGE coefficient functions  $\beta _g, \beta _\lambda ,
 \gamma , \gamma _m$
\etc\  just give enough information. This point was already explained
in detail in the previous paper.\refmark{\BKMNI}
Thus, as far as the region
$\region$ is concerned,
{\it with $L$-loop \ep\ and $(L+1)$-loop RGE coefficient functions,
we can obtain an RGE improved \ep\
which is exact up to $L$-th-to-leading log order.}

Before explaining how to obtain the \ep\ valid in the
complementary region $\cregion$, let us give an explicit expression of
the \ep\ which is obtained by the procedure up to here and
exact in the leading log order in the region $\region$.

The one-loop \ep\ $V_1 = V^{(0)} + V^{(1)}$
is given by the usual formula\refmark{\CW} in the \MS scheme as
$$
V_1 = {m^2\over 2}\phi ^2 +{\lambda \over 4!}\phi ^4 + hm^4 +
{1\over 64\pi ^2}\bigg[ \MB^4\Big(\ln{\MB^2\over \mu ^2}-{3\over 2}
\Big)
 -4N\MF^4\Big(\ln{\MF^2\over \mu ^2}-{3\over 2}\Big) \bigg] \ .
\eqn\eqoneloop
$$
The coefficient functions in RGE are found at the one-loop order as
$$
\eqalign{
\beta _\lambda  &= {1\over 16\pi ^2}(3\lambda ^2+8N\lambda g^2-48Ng^4)
   \equiv \beta _{\lambda 1}\lambda ^2+\beta _{\lambda g}\lambda g^2+
\beta _{\lambda gg}g^4 \ ,\cr
\beta _g &= {1\over 16\pi ^2}(4N+6)g^4 \equiv  \beta _{g1}g^4 \ ,\cr
\gamma _m &= -{1\over 16\pi ^2}(\lambda +4Ng^2) \equiv  \gamma _
{m\lambda 1}\lambda +\gamma _{mg1}g^2 \ ,\cr
\gamma  &= {1\over 16\pi ^2}2Ng^2 \equiv  \gamma _1g^2 \ ,\cr
\beta _h &= +2h\gamma _m +{1\over 16\pi ^2}{1\over 2} \equiv   2h
\gamma _m +\beta _{h1} \ .\cr
}\eqn\eqBG
$$
If one calculates only the anomalous dimension $\gamma $ among these,
all the others can in fact be found immediately
by substituting the one-loop \ep\  \eqoneloop\
into the RGE \eqRGE\ itself.

The barred quantities in the solution \eqSOLTWO\ are obtained in the
same way
as described in the previous paper:\refmark{\BKMNI} That is, we
change the
variable from $t$ to $\widebar s\equiv \widebar s(t)$ in the
RG running
equations $d\widebar g^2/dt = \beta _g(\widebar g^2, \widebar
\lambda )$, \etc,
and integrate them from $\widebar s=s$ to $\widebar s=0$. Using the
$\beta $ and $\gamma $
functions in \eqBG, we find
$$
\eqalign{
&\widebar g^2 = g^2\Big( 1-{\beta _{g1}\over 2}s \Big)^{-1}\ , \qquad
\widebar\phi  = \phi \Big( 1-{\beta _{g1}\over 2}s \Big)^{\gamma _1
\over \beta _{g1}} \ , \cr
&\widebar\lambda  =  g^2\, {
 a(\lambda -bg^2)(g^2/\widebar g^2)^{a{\beta _{\lambda 1}\over \beta _
{g1}}-1} -
    b(\lambda -ag^2)(g^2/\widebar g^2)^{b{\beta _{\lambda 1}\over
\beta _{g1}}-1}  \over
 (\lambda -bg^2)(g^2/\widebar g^2)^{a{\beta _{\lambda 1}\over \beta _
{g1}}} -
    (\lambda -ag^2)(g^2/\widebar g^2)^{b{\beta _{\lambda 1}\over
\beta _{g1}}}  } \ , \cr
&\widebar m^2 =  m^2 \Big({g^2\over \widebar g^2}\Big)^{\gamma _{mg1}
\over \beta _{g1}}
\left(    { (\lambda -bg^2)(g^2/\widebar g^2)^{a{\beta _{\lambda 1}
\over \beta _{g1}}} -
      (\lambda -ag^2)(g^2/\widebar g^2)^{b{\beta _{\lambda 1}\over
\beta _{g1}}} \over (a-b)g^2 }
         \right)^{\gamma _{m\lambda 1}\over \beta _{\lambda 1}} \ ,\cr
&\widebar h\widebar m^4 =  h m^4
   + {m^4\over g^2}{\beta _{h1}\over \beta _{g1}}
   F\Big( {\widebar g^2 \over g^2},\, {\lambda \over g^2} \Big) \ ,
\cr
}\eqn\eqbar
$$
where $a$ and $b$ are the two roots of quadratic equation
$\beta _{\lambda 1}y^2 + (\beta _{\lambda g}-\beta _{g1})y + \beta _
{\lambda gg}=0$,
and the function $F$ is defined by
$$
F(x,\,y) \equiv   \int \NL_1^x dt\  t^{-2({\gamma _{mg1}\over
\beta _{g1}}+1)}
\left( {  a-b  \over
 (y-b)\,t^{-a{\beta _{\lambda 1}\over \beta _{g1}}} -
  (y-a)\,t^{-b{\beta _{\lambda 1}\over \beta _{g1}}} } \right)^{-2
{\gamma _{m\lambda 1}\over \beta _{\lambda 1}}}\ .
\ee
$$

As the `boundary' function of our RGE solution \eqSOLTWO,
we use the one-loop \ep\ $V_1$ at $s=0$
which is given simply by setting $\mu ^2=\fm$
{\it directly} in Eq.\eqoneloop:
$$
\eqalign{
V_1\big\vert_{s=0} = &{m^2\over 2}\phi ^2 +{\lambda \over 4!}\phi ^4 +
 hm^4 \cr
 &+ {1\over 64\pi ^2}\bigg[
 (\sm)^2\Big(\ln{\sm\over \fm }-{3\over 2}\Big) + 6N(\fm)^2 \bigg] \ .
\cr
}\eqn\eqBDF
$$
To the leading-log order, the tree potential part $V^{(0)}\big
\vert_{s=0}$ is
enough. But, as explained in the previous paper,\refmark{\BKMNI}
retaining the one-loop part
$V^{(1)}\big\vert_{s=0}$ makes the approximation better also in the
region in which the log-factor $\ln(\MF^2/\mu ^2)$ is not so large.
Replacing all the parameters $g^2, \lambda , m^2, \phi $ and $hm^4$
in \eqBDF\
by the above obtained barred ones \eqbar, Eq.\eqBDF\ gives the desired
\ep\ which is leading-log `exact' in the region $\region$.

Up to here the procedure is essentially the same as in the previous
paper.
Now we turn to our main task of this paper to develop a method for
obtaining
the \ep\ valid in the complementary region $\cregion$.
For that purpose, we should first recall what we have done in the
above.
We had two logarithm factors
$g^2\ln(\fm/\mu ^2)$ and $g^2\ln[(\sm)/\mu ^2]$. We have chosen the
first factor
as the variable $s$ with which we summed up the leading,
next-to-leading, $\cdots $, log terms, and treated the second factor
also
essentially as $s$ by rewriting it into
$$
g^2\ln{\sm\over \mu ^2} = s + u \ .
\eqn\eqSU
$$
This is all right in the region $\region$ since $u \sim  O(g^2)$ there,
but becomes of course problematic for $g^2\phi ^2 \ll  m^2$ in which
$u$ becomes
very large $\sim O(1)$.

How can we calculate the \ep, or the `boundary' function, in the
region $\cregion$?  The key to this question is to note that it is the
low-energy region. Physically speaking, any heavy
particle, here $\phi $ with mass $m$,
must have decoupled already in such a low-energy
region and the running of the parameters such as couplings, masses
and so on
should be governed solely by the effective low-energy theory containing
no heavy particles. Namely all the heavy particle loop contributions
can be hidden in the redefinition of the low-energy theory parameters.
This is the wisdom of effective field theory
approach.\refmark{\KY-\Georgi}
If we do so,
we have only one mass scale $M_\F^2=\fm$ in the low-energy theory and
so will
not encounter such a problematic variable like $u=\ln[(\sm)/\fm]$,
which
appeared owing to the presence of two different mass scales.

Let us spell out about this in some different way. In the low-energy
region
$\cregion$, the above rewriting \eqSU\ of the second log factor into
$s+u$
is clearly inadequate. Instead, the following expansion
in $\phi ^2/m^2$
becomes good:
$$
g^2\ln{\sm\over \mu ^2} =g^2\Big[\ \ln{m^2\over \mu ^2} + \2{y}{g^2
\phi ^2\over m^2}
    - {y^2\over 8}\Big({g^2\phi ^2\over m^2}\Big)^2 + \cdots \Big] \ .
\eqn\eqEXPAND
$$
In the region $\cregion$, all the terms here except the first one
$g^2\ln(m^2/\mu ^2)$ are of $O(g^2)$
(because of the $g^2$ factor in front) in the sense of leading-log
series
expansion. Note that we have the same form of logarithm expansion as
Eqs.\eqLOGEXP\ and \eqLLOG\
even if we take the second log factor \eqSU\ itself
as the variable $u$. Therefore now the log-factors which we have to
take
account of are $g^2\ln(m^2/\mu ^2)$ and $s=g^2\ln(g^2\phi ^2/\mu ^2)$.
If we must treat these two log factors simultaneously, we would still
have essentially the same difficulty as before. But fortunately,
the decoupling theorem,\refmark{\AC}
or more basically the renormalization theory itself,
guarantees that all the powers of the former log factor, $[\ln(m^2/
\mu ^2)]^p$,
like any positive power terms in $m^2$, can be absorbed into
redefinitions of
the coupling constants and mass parameters in the low-energy
effective field
theory. So if we use those low-energy parameters, the
explicitly appearing log-factor is only $s$, and hence
we can apply our original method for improving the \ep\ with
no problems.
This also explains the point that this approximation is valid only
in the
region $\cregion$ and so really complementary to the previous method:
when
$g^2\phi ^2$ becomes large and comparable as $m^2$, the second and
higher terms
in the expansion \eqEXPAND\ become non-negligible and make
the original
log-factor again, which should be summed up equally as $s$ when
$g^2\phi ^2\gg m^2$.

However, a point may still seem to remain unclear:
if we use the low-energy effective field
theory for the region $\cregion$ while keeping to use original
theory in
$\region$, then what is the relation between the two theories?  In
particular, in this context of RGE improvement of the \ep, what is the
relation between the RGE's of the two theories?

Fortunately this has a very simple and natural answer: {\it The
differential operator ${\cal D}$ in the RGE is in fact unique};
namely the
apparently different  ${\cal D}$ operators in both theories are
the same!

In order to show this explicitly, let us come back to our Yukawa
theory.
There the heavy particle is $\phi $ with mass $m$. The $\phi $'s
one-loop contribution
to the \ep\ was given before as
$ (\MB^4/64\pi ^2)[\ln(\MB^2/\mu ^2)-(3/2)]$ with $\MB^2=\sm$,
which is expanded in $\lambda \phi ^2/m^2$ to yield
$$
 {1\over 64\pi ^2} \bigg[
 m^4\Big(\ln{m^2\over \mu ^2}-{3\over 2}\Big)
   + \lambda m^2\phi ^2\Big(\ln{m^2\over \mu ^2}-1\Big)
   + {\lambda ^2\phi ^4\over 4}\ln{m^2\over \mu ^2}\bigg]\
      + \ O\Big({\lambda \phi ^2\over m^2}\Big)\times \phi ^4 \ .
\eqn\eqHAJIME
$$
As expected all the effects of the heavy particle-loop are to shift the
low-energy theory parameters, the vacuum energy $hm^4$, mass parameter
 $m^2$
and coupling constant $\lambda $, aside from the non-renormalizable
type higher
power terms in $\phi $ all of which are suppressed by  powers of
$\phi ^2/m^2$.
This thus implies that the low-energy theory has the following
mass $\tm$, coupling $\tl$ and vacuum-energy $\th\tm^4$ parameters:
$$
\eqalign{
\tl &= \lambda  + {1\over 16\pi ^2}\2{3\lambda ^2}\ln{m^2\over \mu ^2}
 \ ,\cr
\tm^2 &= m^2 + {1\over 16\pi ^2}\2{\lambda m^2}\Big(\ln{m^2\over
\mu ^2}-1\Big) \ ,\cr
\th\tm^4 &= hm^4 + {1\over 16\pi ^2}{m^4\over 4}\Big(\ln{m^2\over
\mu ^2}-\2{3}\Big) \ .\cr
}\eqn\eqRELATION
$$
These are of course relations valid at one-loop level, and the
higher-loop
corrections give contributions of higher power terms in
$\ln(m^2/\mu ^2)$ .
If we were discussing {\it effective action }  $\Gamma [\phi ,\psi ,
\widebar \psi ]$
instead of \ep, which contains Yukawa term
$-g\widebar\psi  \phi \psi $ in the
tree part, we would find that Yukawa coupling $g$ is also shifted as
\foot{Even if we discuss only the effective potential, we can find this
shift (25) of the Yukawa coupling $g$. This is found, however,
by computing
{\it two-loop} contributions since $g$ appears in the \ep\ only from
the one-loop level. }
$$
\tg^2 = g^2  + {1\over 16\pi ^2}g^4\Big( 3\ln{m^2\over \mu ^2}
-{5\over 2}\Big) \ .
\eqn\eqADD
$$
The scalar field $\phi $ remains the same, \ie, $\tphi=\phi $, as a
special situation
at the one-loop order of this model; the $\phi $'s one-loop diagram
does not
contribute to the wave-function renormalization of $\phi $.
We now rewrite the RG differential operator ${\cal D}$ in \eqRGED\ in
terms
of these new parameters of the low-energy theory:
$$
\eqalign{
\calD &= (\calD\mu )\der\mu +(\calD\tl)\der\tl+(\calD{\tg^2})
\der{\tg^2}
 +(\calD\tm^2)\der{\tm^2}+(\calD\tphi)\der{\tphi}+(\calD\th)\der{\th}
 \cr
 &= \mu \der\mu +{\widetilde \beta _\lambda }\der{\tl}+{\widetilde
\beta _g}\der{\tg^2}
   -{\widetilde \gamma _m}\tm^2\der{\tm^2}-{\widetilde \gamma }\tphi
\der{\tphi}
   +{\widetilde \beta _h}\der{\th} \ ,\cr
}\ee
$$
where using the one-loop relations \eqRELATION\ and \eqADD\  we have
$$
\eqalign{
{\widetilde \beta _\lambda }&= \calD\tl
   = {1\over 16\pi ^2}(8N\lambda g^2 - 48Ng^4) + O(\hbar^2)  \ ,\cr
{\widetilde \beta _g}&= \calD\tg^2
   = {1\over 16\pi ^2}4Ng^4 + O(\hbar^2)  \ ,\cr
{\widetilde \gamma _m}&= -\calD\,\ln\tm^2
   = -{1\over 16\pi ^2}4Ng^2 + O(\hbar^2)  \ ,\cr
{\widetilde \beta _h}&= -\calD\th
   = 2h{\widetilde \gamma _m} + O(\hbar^2)  \ ,\cr
}\eqn\eqTILDE
$$
and ${\widetilde \gamma }=\gamma =2Ng^2/16\pi ^2 + O(\hbar^2)$.
Note that one can freely replace the parameters here
in the RHS's of \eqTILDE\ by the tilded ones since the difference is
of
$O(\hbar^2)$.
We immediately notice here that these are nothing but
the $\beta $ and $\gamma $ functions in the low-energy effective
field theory
in which the running is governed solely by the light particle
(here $\psi $) loop effects, as is clearly seen from the fact that
they are all
now proportional to $N$, the number of fermion species.
This results may sound as a matter of course. But the
important recognition here is that the renormalization group $\calD$
operator
is the same one between the low-energy effective theory and
the original theory. So even if we solve the RGE in the low-energy
effective
field theory, it is guaranteed that we are solving the {\it same}
RGE as in
the original theory simply by using different set of parameters.
Therefore
it is also trivial that the solutions obtained in those two ways
agree with
each other at least around $g^2\phi ^2 \sim  m^2$
where the approximations adopted in
the two methods are both valid.  The parameters should be matched
via the
relations like \eqRELATION. Since the relations contain powers of
the log factor $(\lambda \ {\rm or}\ g^2)\ln(m^2/\mu ^2)$, the
parameter matching
between the low energy theory and the original theory {\it has to}
be done at
a renormalization point $\mu $ around $\mu \sim m$. Otherwise the
unknown higher loop
corrections may become large.

We are now ready to demonstrate the procedure for obtaining the
\ep\ in the complementary region $\cregion$ by explicit
computations to the leading-log order.
Now the Yukawa coupling is $\tg$,
so we use $\ts\equiv \tg^2\ln (\tg^2\tphi/\mu ^2)$ in place of $s$,
and the one-loop potential $V_1$ in \eqoneloop\ with $\MF$ replaced by
$ \tMF\equiv \tg\tphi$,
although the differences are of next-to-leading log\ (or two-loop)
order.
The boundary function is given by the one-loop \ep\  $V_1$ at
$\mu ^2=\tg^2\tphi^2$ (\ie, $\ts=0$).
But, from the $\phi $'s one-loop contribution,
we should subtract the first three terms up to $\phi ^4$ in \eqHAJIME\
since they are absorbed in the redefinitions of the parameters, $h
\rightarrow  \th$,
$m^2 \rightarrow  \tm^2$ and $\lambda \rightarrow  \tl$. The
leading-log \ep\ is obtained by
replacing the parameters there by the barred ones and so is given by
$$
V = {1\over 2}\widebar{\tm}^2 \widebar\tphi^2
   +{1\over 4!}\widebar{\tl}\widebar\tphi^4
   + \widebar{\th}\widebar{\tm}^4
    + {1\over 64\pi ^2}\Big[
\widebar\tm^4\,G\Big({\widebar\tl\widebar\tphi^2\over 2\widebar\tm^2}
\Big)
   + 6N(\widebar \tg^2 \widebar\tphi^2)^2 \Big] \ .
\eqn\eqSOLLAST
$$
where the $G$ term denotes the rest contribution (higher than
$\phi ^4$) of the
$\phi $-loop,
$$
G(x) = (1+x)^2\ln(1+x) - x -{3\over 2}x^2\ .
\ee
$$
Although this contribution is at most of $O(g^2)$ in the region
$\cregion$,
we retained it since it makes better the matching of the present \ep\
with the previous one around  $g^2\phi ^2\sim m^2$.

Since the RG running equation for barred quantities
was already solved in the full theory before,
we can find the solution in this case simply by substituting
$\beta _{\lambda 1}=\gamma _{m\lambda 1}=\beta _{h1}=0$ and
$\beta _{g1}=4N/16\pi ^2\equiv \tbeta$ in the previous
solutions. Therefore the barred quantities in our potential \eqSOLLAST\
are found to be
$$
\eqalign{
&\widebar \tg^2 = \tg^2\Big( 1-{\tbeta\over 2}\ts \Big)^{-1} \ ,
\qquad
\widebar\tphi = \tphi\Big( 1-{\tbeta\over 2}\ts \Big)^{\gamma _1\over
\tbeta} \ ,\cr
&\widebar{\tl} = (\tl - a\tg^2)\Big( 1-{\tbeta\over 2}\ts \Big)^{
-{\beta _{\lambda g}\over \tbeta}}
   \qquad \Big( a\equiv {\beta _{\lambda gg}\over \tbeta-\beta _
{\lambda g}}\Big) \ ,\cr
&\widebar{\tm}^2 = \tm^2 \Big( 1-{\tbeta\over 2}\ts \Big)^{{\gamma
_{mg1}\over \tbeta}} \ ,
\qquad
\widebar{\th}\widebar{\tm}^4 = \th\tm^4 \ .\cr
}\ee
$$
Now we match the (unbarred) tilded parameters (which are also running
and
functions of renormalization point $\mu $) with the untilded ones
by choosing
the renormalization point $\mu =m$. Then the relations
\eqRELATION\ and \eqADD\  give\  (aside from $\tphi=\phi $)
$$
\eqalign{
\tl &= \lambda \ , \quad \qquad   \tg^2 = g^2 - {1\over 16\pi ^2}
{5g^4\over 2} \ ,\cr
\tm^2 &= m^2 - {1\over 16\pi ^2}\2{\lambda m^2} \ , \qquad
\th\tm^4 = hm^4 - {1\over 16\pi ^2}{3m^4\over 8} \ .\cr
}\eqn\eqRELAT
$$

The agreement of  this \ep\ \eqSOLLAST\
by low-energy effective theory with that
obtained previously in the original theory in the region $g\phi
\sim m$ will be
clear from the derivation. Since the both potentials satisfy the RGE
they
are $\mu $-independent ($dV/d\mu =0$).
So we can compare them choosing $\mu =m$.
Then at $(g\phi \simeq\tg\tphi)=m$ the parameters $s\simeq\ts$ equal
zero and
hence all the barred quantities reduce to the unbarred ones
in both expressions. Namely,
the both potentials reduce to the ``boundary functions" in each scheme.
But they coincide with each other\ (up to two-loop quantities) by
construction under identification \eqRELAT.

We have explained our procedure by using the simplest example of
Yukawa model.
However the method described here is quite general and indeed
applicable to
any complicated systems. We conclude this paper by adding some
explanations
how we can improve the \ep\ of multi-scalar fields applying the above
procedure. For illustration let us consider the case of
the two scalar field potential $V(\phi _1, \phi _2)$,
\foot{Our procedure described here should be compared with a more
complicated
method proposed by Einhorn and Jones\refmark{\EJ} in which they
introduce two renormalization points for $\phi _1$ and $\phi _2$.}
in a general system
consisting of many particles which couple to those two scalar fields.
In such a system, we would have typically the following three types of
logarithm factors:
$$
s_1 \equiv  g_1^2 \ln{g_1^2\phi _1^2\over \mu ^2}\ , \quad
s_2 \equiv  g_2^2 \ln{g_2^2\phi _2^2\over \mu ^2}\ , \quad
s_3 \equiv  \lambda _3 \ln{\lambda _1\phi _1^2 + \lambda _2\phi _2^2
+m^2 \over \mu ^2}\ ,
\ee
$$
where $s_1$- and $s_2$-type factors come from fermion loops
and $s_3$-type from scalar loops.\foot{
Here we are not claiming complete generality. We are just trying to
explain a
typical procedure which will be applicable to more general systems.
}
[The minimal supersymmetric standard model is similar to this example,
 where
$\phi _1$ and $\phi _2$ correspond to the two Higgs doublets and $m$
to the
supersymmetry breaking scale.]
Let us call the particles which produce the $s_i$-type log factors the
type-$i$ particles.
We consider in any case that the coupling constants are of the same
order,
$g_1^2 \sim g_2^2 \sim \lambda _1 \sim \lambda _2 \sim  \lambda _3$.

To treat the $s_1$- and $s_2$-type log
factors well, we first have to separate the $(\phi _1^2, \phi _2^2)$
plane into
two regions, $\phi _1^2\siml\phi _2^2$ and $\phi _1^2\simg\phi _2^2$,
and derive the \ep\
in the two regions separately.  Now consider the first region
$\phi _1^2\siml\phi _2^2$.  (Second region is obtained in the same
way by
exchanging the role of $\phi _1$ and $\phi _2$.)
Then we may regard
the two scalar field potential $V(\phi _1, \phi _2)$ as a single
scalar field
potential $V_{\phi _2}(\phi _1)$ of $\phi _1$ with a {\it parameter}
$\phi _2$.
There the type-2 and type-3 particles can be viewed as particles
carrying ($\phi _2$-dependent) masses $g_2^2\phi _2^2$ and
$\lambda _2\phi _2^2+m^2$,
respectively.  Both of them are heavy in this region
$\phi _1^2\siml\phi _2^2$,
and therefore decouple.  So, in this region, the log factors which we
have
to treat explicitly in the \ep\ are only $s_1$-type. The $s_2$ and
$s_3$
log factors are taken into account simply by redefining the coupling
constants (and masses) in the low energy theory.  A care, however,
may be
necessary in this redefinition: if $\lambda _2\phi _2^2 \ll  m^2$,
the redefinition
cannot be done in a single step. If we would do it in one step,
the relation
between an original coupling constant, write $\lambda $ generically,
and its counterpart
$\tl$ in the low energy effective theory would take the form like
$$
\tl(\mu ) = \lambda (\mu ) + c_2(\lambda )\ln{g_2^2\phi _2^2\over
\mu ^2} +
           c_3(\lambda )\ln{\lambda _2\phi _2^2+m^2\over \mu ^2}
\ee$$
at the one-loop level with certain coefficients $c_2$ and $c_3$
quadratic in
the coupling constants $\lambda $. The last log
comes from the expansion of the $s_3$-type log factors and the second
log
from $s_2$. The higher loop contributions give higher power terms of
these two log factors. If $\lambda _2\phi _2^2 \ll  m^2$,  then
$g_2^2\phi _2^2 \ll  \lambda _2\phi _2^2+m^2$, so that those two log
factors cannot be made
small simultaneously whatever renormalization point $\mu $ is chosen.
 This
means that we can find no reliable relation between $\lambda $ and
$\tl$ unless we
calculate all the higher loop contributions. Of course we know how
 to avoid
this difficulty. As in the usual treatment of effective field theory
in the
presence of multi-threshold, we should do the redefinition
in two steps:
First, as we come down to the scale $\mu ^2 \sim  \lambda _2\phi _2^2
+m^2$,
we switch to an intermediate energy
effective field theory in which only the type-3 particles decouple
and the
coupling $\lambda $ is shifted to
$$
\tl(\mu ) = \lambda (\mu ) +  c_3(\lambda )\ln{\lambda _2\phi _2^2
+m^2\over \mu ^2}\ .
\ee$$
Next, at the scale $\mu ^2 \sim  g_2^2\phi _2^2$,
we switch to the low energy effective
field theory in which the type-2 particles also decouple and the
coupling
$\tl$ is shifted into
$$
\widetilde{\tl}(\mu ) = \tl(\mu ) + c_2(\tl)\ln{\tg_2^2\tphi_2^2\over
\mu ^2}\ .
\ee$$
Then we have a single log factor in each step of the
coupling redefinition and
so can find a reliable connection condition of the coupling constants.
 At the
final stage, the \ep\ is written in terms of the $\widetilde{\tl}$
 coupling
constants and contain only the $s_1$-type log factors explicitly,
so that it can easily be improved by RGE.

\refout

\bye